\documentclass[twocolumn]{revtex4}
\usepackage{graphicx}

\begin{document}

\title{Effect of the tetragonal distortion on the electronic structure, phonons and superconductivity in the Mo$_{3}$Sb$_7$ superconductor}

\author{Bartlomiej Wiendlocha}
\email[email: ]{wiendlocha@fis.agh.edu.pl}
\affiliation{AGH University of Science and Technology, Faculty of Physics and Applied Computer Science, Al. Mickiewicza 30, 30-059 Krakow, Poland}

\author{Malgorzata Sternik}
\affiliation{Institute of Nuclear Physics, Polish Academy of
Sciences, Radzikowskiego 152, 31-342 Krakow, Poland}

\date{\today}

\title{Effect of the tetragonal distortion on the electronic structure, phonons and superconductivity in the Mo$_{3}$Sb$_7$ superconductor}

\begin{abstract}
Effect of tetragonal distortion on the electronic structure, dynamical properties and superconductivity in 
Mo$_3$Sb$_7$ is analyzed using first principles electronic structure and phonon calculations. 
Rigid muffin tin approximation (RMTA) and McMillan formulas are used to calculate the electron-phonon 
coupling constant $\lambda$ and superconducting critical temperature. Our results show, that tetragonal distortion has small, 
but beneficial effect on superconductivity, slightly increasing $\lambda$, and the conclusion that the electron-phonon mechanism
is responsible for the superconductivity in Mo$_3$Sb$_7$ is supported.
The spin-polarized calculations for the ordered (ferromagnetic or antiferromagnetic), as well as disordered (disordered local moment) magnetic 
states yielded non-magnetic ground state.
We point out that due to its experimentally observed magnetic properties the tetragonal Mo$_3$Sb$_7$ might be 
treated as noncentrosymmetric superconductor, which could have influence for the pairing symmetry. 
In this context the relativistic band structure is calculated and spin-orbit interaction effects are discussed.
\end{abstract}

\keywords{superconductivity, electronic structure, phonons,
electron-phonon coupling}

\maketitle

\section{Introduction}

A Zintl-type intermetallic compound Mo$_3$Sb$_7$ has gained much attention in recent years due to its
interesting superconducting and transport properties. It is a type II
superconductor,\cite{Buk02,Buk06} with the relatively low critical temperature
$T_c\simeq 2.2$~K. At room temperature, it has a cubic {\it bcc} structure (space
group {\it Im3m}) of the Ir$_3$Ge$_7$ type. The primitive cell of
Mo$_3$Sb$_7$ contains two formula units, i.e. 20 atoms, occupying
three nonequivalent positions, listed in Table~\ref{tab-cryst}.
Mo$_3$Sb$_7$ is a system, where superconducting, structural and magnetic properties interpenetrate each other. 
The considerable interest on Mo$_3$Sb$_7$ started, when Candolfi {\it et al.},\cite{cc-prl} suggested that spin fluctuations compete with superconductivity in this compound, modifying also the temperature dependence of electrical resistivity and magnetic susceptibility. 
Later on, Tran {\it et al.} \cite{Tra08} observed a peak in the specific heat $C_P(T)$ at $T^*=50$~K and reanalyzed anomalies in 
magnetization and resistivity, arguing the formation of the spin-singlet dimers and opening of the gap\cite{Tra08,tran-gap-evidence} in spin excitations around $T^*=50$~K. Next, Koyama {\it et al.}\cite{koyama} observed a structural cubic-to-tetragonal distortion below $T^*=53$~K, and suggested the formation of the spin-singlet dimers valence-bond crystal.\cite{koyama2,okabe}

\begin{figure}[htb]
\includegraphics[width=0.48\textwidth]{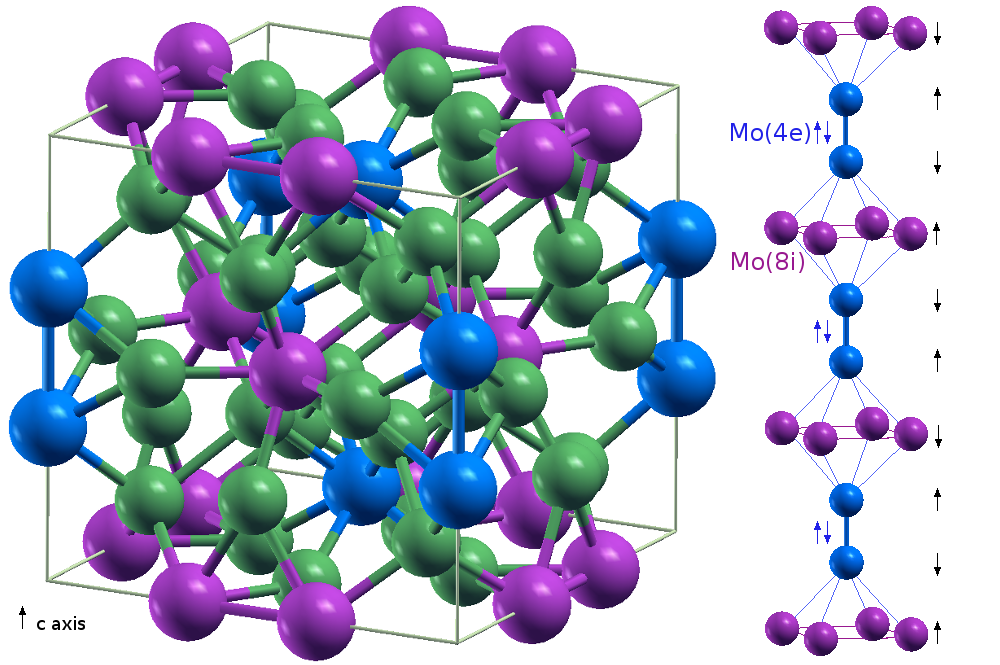}
\caption{\label{fig-cryst}  Left: Conventional unit cell of the tetragonal Mo$_3$Sb$_7$. In blue: Mo(4e), in pink Mo(8i), in green all of the Sb atoms (Sb atoms were made smaller). Right: sublattice of Mo atoms. The singlet dimers formation between nearest Mo(4e) atoms, as indicated by the blue arrows, was proposed,\cite{Tra08,koyama,koyama2,okabe} but the contribution from the Mo(8i) atom should be important to destabilize the cubic phase.\cite{koyama} Black arrows on the right indicate one of the possibilities of static magnetic ordering, considered in our calculations. If the two Mo(4e) atoms, which form singlet dimer, are treated as non-equivalent, the crystal looses inversion center (see, Sec.~\ref{sec2}).}
\end{figure}

As far as the superconducting pairing mechanism is concerned, there is no agreement whether Mo$_3$Sb$_7$ is a typical s-wave electron-phonon (BCS) superconductor or some more complex state is realized. Most of the published results~\cite{cc-prl,Buk06,tran-heat,cc-heat,bw08}, including first principles calculations for the cubic phase,\cite{bw08} support the conventional picture.
However, various values of the electronic specific heat jump at the transition point were reported,\cite{cc-prl,tran-heat,yan2013} $\Delta C/\gamma T_c = 1.04 - 1.56$ much smaller, or close to the weak--coupling BCS value 1.43. Moreover, the specific heat in the superconducting state\cite{cc-heat,tran-heat} and the magnetic field penetration depth\cite{usr-khasanov,usr-tran} behaviors were differently analyzed, in terms of either one-- or two--gap BCS models. 
It was also proposed, that antiferromagnetic fluctuations may be responsible for the superconductivity, due to observation of the spin density wave under pressure.\cite{tran-pressure} 

Recently, basing on the neutron single crystal and X-ray powder diffraction studies, structural parameters of the low temperature tetragonal phase were accurately determined.\cite{okabe,yan2013}
Availability of the exact crystallographic data and renew interest in this system prompted us to re-investigate the electronic structure, magnetism, dynamics and the electron-phonon interaction in the tetragonal Mo$_3$Sb$_7$ to discuss the effect of the distortion on Mo$_3$Sb$_7$ properties. 
Especially we wanted to verify, whether the strength of the electron-phonon interaction in the tetragonal phase is sufficient to drive the system into the superconducting state, with the experimentally observed critical temperature, as the previous results supporting electron-phonon superconductivity\cite{bw08} were obtained for the cubic phase. Due to the distinct crystal structures, it is possible that in the tetragonal phase (where in fact superconductivity in Mo$_3$Sb$_7$ appears) the situation could be much different.\footnote{Note, that Ref. \cite{bw08} was published before the tetragonal distortion was detected.}

Analysis of the unit cell symmetry in terms of the possible formation of antiferromagnetic phases (see, Sec.~\ref{sec2}) led to observation, that when the two Mo(4e) atoms, expected\cite{Tra08,koyama,koyama2,okabe} to form the spin singlet dimers, are taken as nonequivalent, the unit cell looses the inversion symmetry. Since this requires the appearance of the distinguishing physical feature, like anti-parallel direction of local magnetic moment, this is rather not a case of the electronic spin-singlet state. However, under external pressure the antiferromagnetic spin density wave state was suggested to stabilize,\cite{tran-pressure} and this could provide the distinguishing feature between the two Mo(4e) atoms. That could transform Mo$_3$Sb$_7$ into the non-centrosymmetric superconductor and this new feature of the system is briefly discussed in connection with the spin-orbit coupling effects.

\begin{table}[htb]
\caption{Measured and calculated structural parameters for Mo$_3$Sb$_7$ for the tetragonal ($I4/mmm$, no. 139) and cubic ($Im$-$3m$, no. 225) phases. Experimental data are taken from neutron studies (Ref.~\cite{yan2013}). Calculated values are presented in parenthesis. The lattice parameters are a = 9.543 (9.640)~\AA \  for the cubic structure; a = 9.551 (9.641)~\AA , c = 9.523 (9.638~\AA ), a/c = 1.003 (1.0003) for the tetragonal phase. Here and in the text, atoms are named after sites they occupy.}
\label{tab-cryst}
\begin{tabular}{lcccc}
\hline
 Tetragonal & site & \emph{x} & \emph{y} & \emph{z} \\
\hline
 Mo(8i)  & 8\emph{i} &  0.3436    &  0     &  0     \\
         &           & (0.3420)   &  0     &  0     \\
 Mo(4e)  & 4\emph{e} &  0         &  0     &  0.3442   \\
         &           &  0         &  0     & (0.3424)  \\
 Sb(4d)  &4\emph{d}  &  0         &  1/2      &  1/4    \\
 Sb(8j)  &8\emph{j}  &  0.2501    &  1/2      &  0   \\
         &           & (0.2503)   &  1/2      &  0   \\
 Sb(16m) &16\emph{m} &  0.16232   &  0.16232  &  0.1621  \\
         &           & (0.16082)  & (0.16082) & (0.16091)  \\
\hline
Cubic & site & \emph{x} & \emph{y} & \emph{z} \\
\hline
 Mo(12e) & 12\emph{e}&  0.3434    &  0     &  0     \\
         &           & (0.3421)   &  0     &  0     \\
 Sb(12d) &12\emph{d} &  0.25      &  0     &  0.50  \\
 Sb(16f) &16\emph{f} &  0.16219   &  0.16219    &  0.16219  \\
         &           & (0.16082)  & (0.16082)   & (0.16082)   \\
\hline
\multicolumn{5}{c}{Primitive cell site and population splitting after distortion:}\\
\multicolumn{5}{c}{6 $\times$ Mo(12e) $\rightarrow$ 4 $\times$ Mo(8i) + 2 $\times$ Mo(4e)}\\
\multicolumn{5}{c}{6 $\times$ Sb(12d) $\rightarrow$ 2 $\times$ Sb(4d) + 4 $\times$ Sb(8j)}\\
\hline
\end{tabular}
\end{table}

The paper is organized in four sections. In Section~\ref{sec1} we discuss the strength of electron-phonon interaction basing on the rigid muffin tin approximation (RMTA). Here, semirelativistic results as well as the spherical potential approximation is used in the electronic structure calculations, as it is required by the RMTA. In Section~\ref{sec2} we discuss our efforts to investigate magnetism in the Mo$_3$Sb$_7$, and, in the context of the possible lack of inversion center, spin-orbit effects are discussed. In this part, full potential fully relativistic KKR approach was used. Section~\ref{sec3} gives the summary.

\begin{figure}[htb]
\includegraphics[width=0.45\textwidth]{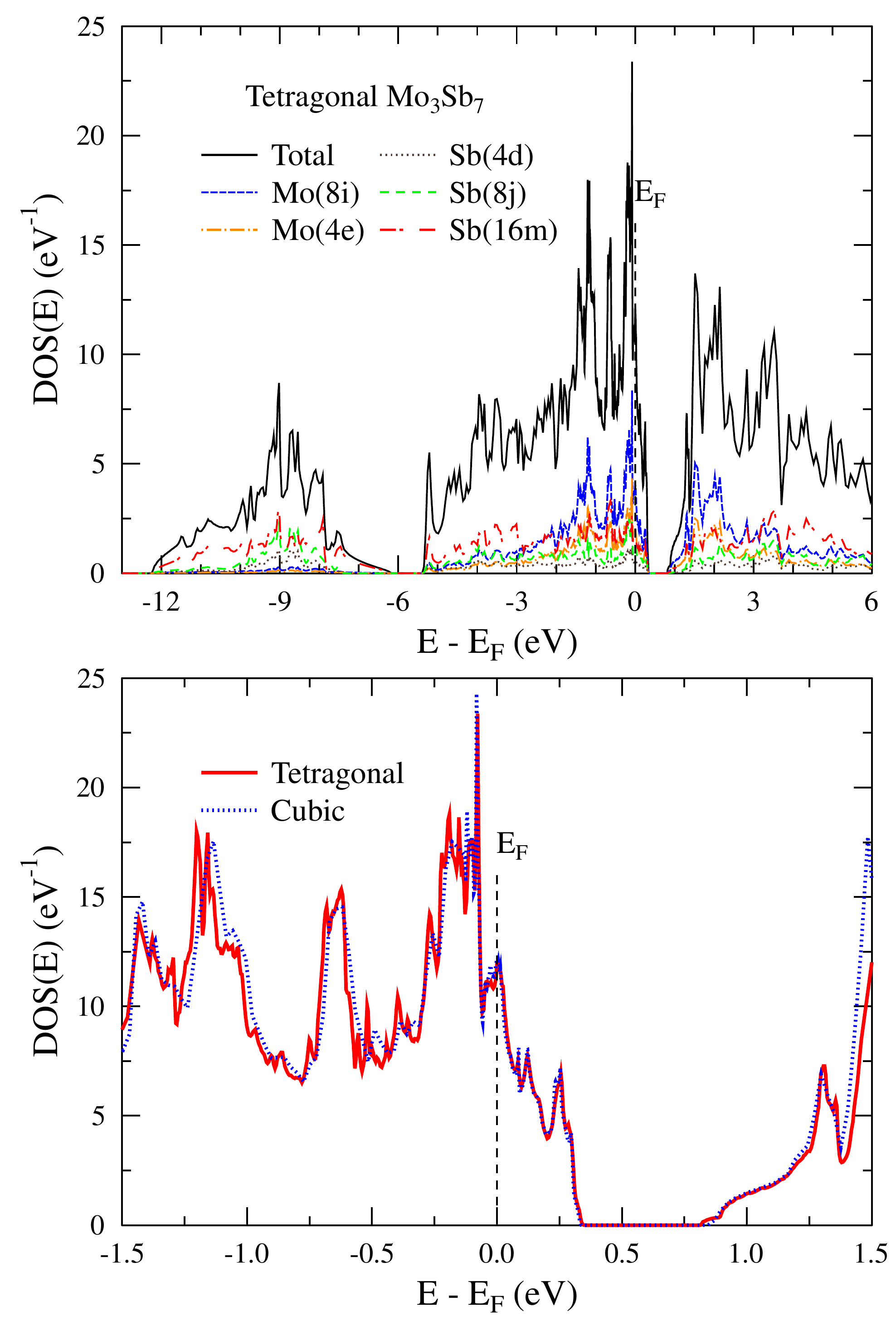}
\caption{\label{fig-tdos}  Top panel: total and site-decomposed
densities of electronic states in Mo$_3$Sb$_7$ (per formula unit).
Bottom panel: comparison of the total DOS near $E_F$ of the cubic and tetragonal phases.}
\end{figure}

\section{Electron-phonon interaction and superconductivity\label{sec1}}

\begin{figure}[htb]
\includegraphics[width=0.45\textwidth]{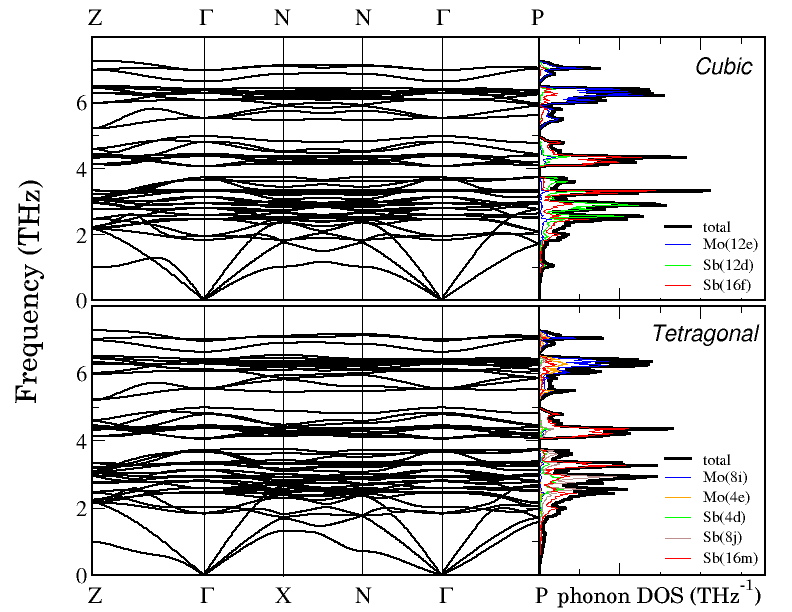}
\caption{\label{fig-phdos}  The phonon dispersions along the high symmetry directions and the total and site-decomposed phonon DOS spectra calculated for cubic (from Ref.\cite{bw08}) and tetragonal Mo$_3$Sb$_7$ structures with lattice parameters and atomic positions optimized in calculations.}
\end{figure}

Electronic structure calculations were performed using the
Korringa-Kohn-Rostoker (KKR) multiple scattering
method.\cite{kkr99} The crystal potential was constructed in the
framework of the local density approximation (LDA), using von
Barth and Hedin formula \cite{lda} for the exchange--correlation
part. 
For all atoms angular momentum cut--off $l_{max}=4$ was set;
\makebox{{\bf k}--point} mesh in the irreducible part of the
Brillouin zone (BZ) contained about 650 points. Density of states
(DOS) was computed using the tetrahedron \makebox{{\bf k}--space}
integration technique, generating about 2600 tetrahedrons in the
irreducible part of the BZ. Since our main goal in this work is to
estimate the electron-phonon coupling constant from first principles 
within the rigid MT approximation, spherical potential approximation for the
crystal potential is used, as is required in this approach.
The muffin tin spheres radii were set to 2.54 $a_B$ (Mo) and 2.73 $a_B$ (Sb), in Bohr atomic units.
Experimental tetragonal unit cell and atomic positions, as shown in Table~\ref{tab-cryst}, were used.
To ensure the consistency of the calculations, electronic structure of the cubic phase was also recalculated using the neutron data of Yan {\it et al.}\cite{yan2013}, which were slightly different from the previously used.\cite{bw08} 
The differences were minor, smaller than those induced by the tetragonal distortion, as can be seen by comparing the data in Ref.~\cite{bw08} and the Table~\ref{tab-dos}. 

Total and site--decomposed electronic densities of states for the tetragonal Mo$_3$Sb$_7$ are
presented in the top panel of Fig.~\ref{fig-tdos}, whereas in the bottom panel, comparison of the 
tetragonal and cubic DOS is shown. As one can see, overall DOS of the tetragonal phase is very similar to the cubic one, which confirms the subtle nature of the distortion. 
The differences are seen on the numerical values of the site-decomposed DOSs and McMillan-Hopfield parameters, presented in Table~\ref{tab-dos}, as well as on the electron dispersion curves, which are discussed in more details in Section~\ref{sec2}.
As in the cubic phase, the characteristic feature of the DOS of tetragonal Mo$_3$Sb$_7$ 
is the presence of a band gap, separating valence and conduction bands, with the Fermi level
$E_F$ located in the range of sharply decreasing valence DOS.
Electronic states near $E_F$ are built
out of the Mo--$4d$ and Sb--$5p$ states, with the highest DOS contribution coming from
Mo atoms. 


It is worth recalling, that the steep DOS function in the vicinity of the band gap also result in interesting thermoelectric properties of Mo$_3$Sb$_7$, studied in considerable number of papers, see, e.g. Refs.~\cite{mosb-transport,mosb-thermo1,mosb-thermo2,mosb-thermo3,mosb-thermo4,mosb-thermo5,mosb-thermo6}.

\begin{table*}[htb]
\caption{Site--decomposed electronic and dynamic properties of
Mo$_3$Sb$_7$. $n_i(E_F)$ is DOS in Ry$^{-1}$/spin, $\eta_i$ in
mRy/${a_B}^2$ (both per atom), $\omega_i$ in THz. Values of
$\lambda_i$, to facilitate the comparison between phases, are also given per one atom, 
and the proper summation with the relative changes in coupling among sublattices (in parenthesis) is done at the bottom of the Table.\label{tab-dos}} 
\begin{tabular}{lccccccccccc}
\hline
atom & $n_i(E_F)$ & $n_s(E_F)$ &$n_p(E_F)$ &$n_d(E_F)$ &
$n_f(E_F)$ & $\eta_i$ &
$\eta_{sp}$ & $\eta_{pd}$ & $\eta_{df}$ & $\sqrt{\langle\omega_i^2\rangle}$  & $\lambda_i$\\
\hline
\multicolumn{12}{c}{Tetragonal}\\
Mo(8i) & 13.80 & 0.051& 0.541 & 13.16 & 0.039  & 6.69 &  0.0 & 1.33 & 5.35 & 5.13  & 0.0315  \\
Mo(4e) & 14.15 & 0.054& 0.533 & 13.50 & 0.044  & 7.48 &  0.0 & 1.34 & 6.13 & 5.07  & 0.0360  \\
Sb(4d) &  3.66 & 0.098& 3.107 & 0.361  & 0.065  & 2.63 &  0.0 & 2.62 & 0.01 & 2.97 & 0.0300  \\
Sb(8j) &  3.71 & 0.092& 2.150 & 0.373  & 0.066  & 2.75 &  0.0 & 2.74 & 0.01 & 2.99 & 0.0305  \\
Sb(16m) &  2.97 & 0.137& 2.147 & 0.498  & 0.165  & 2.59 &  0.0 & 2.56 & 0.03 & 3.42& 0.0212  \\
\hline
\multicolumn{12}{c}{Cubic}\\
Mo(12e) & 14.09 & 0.053& 0.523 & 13.46 & 0.040  & 6.87 &  0.0 & 1.31 & 5.54 & 5.07 & 0.0324  \\
Sb(12d) &  3.59 & 0.088& 3.041 & 0.367 & 0.066  & 2.60 &  0.0 & 2.59 & 0.01 & 2.97 & 0.0282  \\
Sb(16f) &  2.98 & 0.136& 2.155 & 0.504 & 0.169  & 2.63 &  0.0 & 2.59 & 0.04 & 3.41 & 0.0218  \\
\hline
\multicolumn{12}{c}{Electron-phonon coupling parameters modifications after distortion:}\\
\multicolumn{6}{r}{0.194 $\rightarrow$ 0.127 + 0.072 = 0.199 (+2.6\%) } &
\multicolumn{6}{l}{[6$\times$ Mo(12e) $\rightarrow$ 4$\times$ Mo(8i) + 2$\times$ Mo(4e)]}\\

\multicolumn{6}{r}{0.169 $\rightarrow$ 0.060 + 0.122 = 0.182 (+7.7\%)} &
\multicolumn{6}{l}{[6$\times$ Sb(12d) $\rightarrow$ 2$\times$ Sb(4d) + 4$\times$ Sb(8j)]}\\

\multicolumn{6}{r}{0.174 $\rightarrow$ 0.169 (-2.9\%)} &
\multicolumn{6}{l}{[8$\times$ Sb(16f) $\rightarrow$ 8$\times$ Sb(16m)]}\\
\multicolumn{12}{c}{$\lambda_{\rm Cubic} = 0.538 \rightarrow \lambda_{\rm Tetragonal} = 0.550$ (+2.2\%)}\\
\hline
\end{tabular}
\end{table*}

Next, we present the phonon calculations results, as a next step towards the discussion of the electron-phonon interaction strength.
In the phonon calculations, both Mo$_3$Sb$_7$ phases were modeled by imposing the symmetry restrictions of the Im-3m (cubic) and I4/mmm (tetragonal) space groups on the crystal structure. Results for the cubic structure have been already presented in our previous paper. \cite{bw08} The calculations of tetragonal phase have been performed using the same technique. 
Structure optimization was achieved using the VASP package.\cite{vasp}
The spin-polarized density functional total energy calculations were carried out within the generalized gradient approximation and using the Perdew, Burke, and Ernzerhof (PBE) functional.\cite{Per96} 
The wave functions were sampled according to Monkhorst--Pack scheme with a {\bf k}--point mesh of (4,4,4). 
The structural calculations were performed on a $\sqrt{2}\times\sqrt{2}\times 1$ supercell (containing 80 atoms) with periodic boundary conditions. 
The optimization of this system started from the experimental lattice constants and atomic positions. 
The lattice parameters and atomic positions obtained after structure optimization of cubic and tetragonal phases are presented in Table~\ref{tab-cryst}. 
The determined values slightly differ from experimental data. 
The calculated lattice constants are longer than the measured parameters of about 1\%  and the calculated tetragonality of I4/mmm phase is weaker than that obtained experimentally. 
The total energy of both structures are almost equal (difference is 22 meV for 80 atoms supercell) thus the energetically none of them is favored in a low temperature region. 

In the static DFT calculations, the magnetic feature of materials with unknown magnetic ordering are usually approached by the ferro- or antiferromagnetic arrangements of the local moments. Unfortunately, in calculations of Mo$_3$Sb$_7$ the starting non-zero moments aligned in the fero- or antiferomagnetic ordering dropped to zero leading up the system to a nonmagnetic state. The lack of antiferromagnetic interaction, which should shorten both the Mo-Mo distance and the lattice parameter c, is then responsible for the small tetragonality of the theoretical unit cell.

For the optimized structure, the phonon dispersions and DOS were calculated with the direct method.\cite{Par97} 
This method utilizes Hellmann-Feynman forces obtained by performing small atomic displacements of nonequivalent atoms from their equilibrium positions. 
From them the dynamical matrix is determined and diagonalized to obtain the phonon frequencies at each wave vector.  
The phonon dispersion relations and the total and site--decomposed partial phonon DOS spectra computed by random sampling of the BZ  are presented in
Fig.~\ref{fig-phdos} for the cubic (top panel) and tetragonal (bottom) optimized supercells. The representative path between high symmetry points in reciprocal space of a body-centered cubic structure is fixed on H(1/2,1/2,-1/2)--$\Gamma$(0,0,0)--N(0,0,1/2)--N(-1/2,0,1/2)--$\Gamma$(0,0,0)--P(1/4,1/4,1/4). For the tetragonally distorted structure the H point is replaced by Z, and two N points are distinguished giving two different points X and N. The considerable differences observed between the dispersion relations of cubic and tetragonal phases are related to the acoustic branches only.  The frequencies calculated at points X and N of tetragonal phase are slightly higher or lower than the frequency at point N of cubic structure. The lowest degenerated acoustic branches in $\Gamma$--P direction in the cubic phase split to two different branches in the tetragonal phase. 

The phonon DOS spectrum of tetragonal phase looks similar to  the previously calculated\cite{bw08} spectrum for the cubic structure. 
The pronounced difference is the small peak observed at 1~THz for the cubic structure and invisible in a low-frequency region of tetragonal phase spectrum. 
This low energy mode was discussed in a previous paper where the experimentally derived phonon DOS of polycrystalline Mo$_3$Sb$_7$ are analyzed and compared with the calculated data.\cite{GenPDOS} 
We showed that the very small changes of the Mo-Mo force constant parameters are able to renormalize the phonon frequencies around the point H to distinctly higher values. 
The tetragonal deformation gives similar results spreading the phonon frequencies around 1 THz.  Finally, the phonon DOS calculated for the tetragonal phase  better corresponds to the experimental data. 

The electronic structure results were used to calculate the
electronic part of the electron-phonon coupling constant, i.e. the McMillan--Hopfield
$\eta_i$ parameters\cite{mcm,hop} for each atom in both phases, using the
formula:\cite{rmta,pickett}
\begin{equation}
\label{eq:eta} \eta_i =\!\sum_l \frac{(2l + 2)\,n_l\,
n_{l+1}}{(2l+1)(2l+3)N(E_F)} \left|\int_0^{R_{\mathsf{MT}}}\!\!r^2
R_l\frac{dV}{dr}R_{l+1} \right|^2\!,
\end{equation}
where $V(r)$ is the self-consistent potential at site $i$,
$R_\mathsf{MT}$ is the radius of the $i$-th MT sphere, $R_l(r)$ is
a regular solution of the radial Schr\"odinger equation
(normalized to unity inside the MT sphere), $n_l(E_F)$ is the
$l$--th partial DOS per spin at the Fermi level $E_F$, and
$N(E_F)$ is the total DOS per primitive cell and per spin.

The phonon DOS $F(\omega)$ was used to compute the average square site--decomposed phonon frequencies $\langle\omega_i^2\rangle$ presented in the 
Table~\ref{tab-dos}, using the formula:\cite{grimvall}
\begin{equation}
\langle\omega_i^2\rangle = \int_0^{\omega_{\mathsf{max}}} \omega
F_i(\omega) d\omega \left/ \int_0^{\omega_{\mathsf{max}}}
\frac{F_i(\omega)}{\omega}d\omega \right.
\end{equation}
The combination of the McMillan-Hopfield parameters and phonon frequency moments allows to calculate the electron-phonon
coupling parameter, according to the equation:
\begin{equation}
\label{eq:lam} \lambda = \sum_i \frac{\eta_i}{M_i\langle
\omega_i^2 \rangle} = \sum_i \lambda_i.
\end{equation}
The sum is over all the $i$ atoms in the primitive cell, $M_i$
is the atomic mass. 
For a review of previous results and a more detailed discussion of the approximations involved in this approach, see e.g.
Refs.~\cite{pss-bw,prb-bw,cr3gan-bw}.

For each nonequivalent atom in both
cubic, and tetragonal phases, values of
$\eta_i$ parameters are presented in
Table~\ref{tab-dos}, with contributions from each
$l\rightarrow l+1$ channels.
As far as the contributions to $\eta_i$ are concerned, typical tendencies 
for p- and d-like elements can be observed. For Mo (4d element), the $d$--$f$ channel is the most
important, whereas $p$--$d$ contribution dominates for both Sb atoms (5p element). 
The tetragonal distortion changes the values of McMillan--Hopfield parameters in both ways. For the group of Mo(12e) atoms, which is split into Mo(8i) and Mo(4e), $\eta$ decreases for the first and increases for the second. The larger value of $\eta$ for Mo(4e) together with unchanged average phonon frequency makes the contribution of this atom to the total electron-phonon coupling constant higher than for the Mo(8i) atoms, if values per one atom are compared. The accompanying increase of the $\langle\omega^2\rangle$ value for the Mo(8i) and decrease of $\eta$ does not allow for the substantial increase of the overall contribution to EPC constant from the group of Mo atoms, which increase only in about 2.6\%. Tetragonal distortion, through the modifications in the electronic structure, increases the partial $\lambda$ for Sb(12d) group of atoms -- overall contribution from Sb(4d) and Sb(8j) is 7.7\% larger. Finally, for the last Sb(16f) group we observe slight decrease of McMillan--Hopfield parameters, as 
well as increase of the averaged phonon frequency, resulting in 2.9\% decrease of the coupling constant. After summation of the atomic contributions, taking into account the population of atoms among the sites, the total electron-phonon coupling constant in the tetragonal phase, $\lambda_{\rm Tet} = 0.550$, is higher than the corresponding value before the distortion, $\lambda_{\rm Cub} = 0.538$. 

\begin{figure}[htb]
\includegraphics[width=0.45\textwidth]{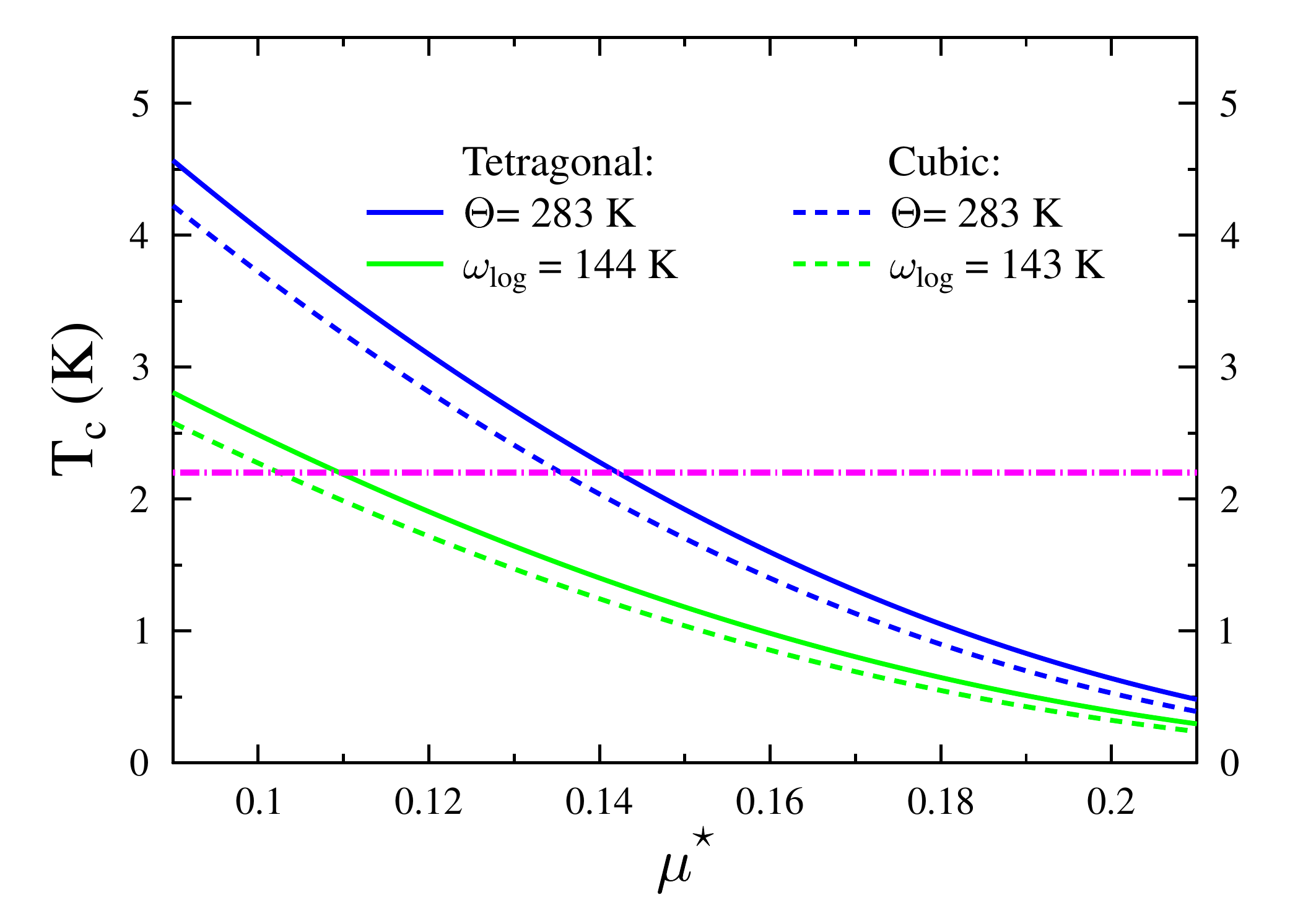}
\caption{\label{fig-tc-mu}  Critical temperature
$T_c$ as a function of Coulomb parameter $\mu^{\star}$ for both phases of Mo$_3$Sb$_7$. 
Curves, marked by $\Theta$, were computed using the original McMillan formula, while for $\omega_{\rm log}$ the modified Allen-Dynes\cite{allen} version was used. Vertical line marks experimental T$_c$ = 2.2~K.}
\end{figure}

As far as the relative importance of the Mo and Sb elements in EPC magnitude is concerned, tetragonal distortion doesn't change the conclusion drawn for the cubic phase.\cite{bw08}
Despite the dominant character of Mo states near $E_F$, electron-phonon interaction has similar strength among all of the atoms. 
The small increase of $\lambda$ leads to the small variation in the superconducting critical temperature $T_c$, which we calculate using McMillan formula:\cite{mcm} 

\begin{equation}\label{eq:tc}
T_c =  \frac{\Theta}{1.45}\,
\exp\left\{-\frac{1.04(1+\lambda)}{\lambda-\mu^{\star}(1+0.62\lambda)}\right\},
\end{equation}
where $\Theta$ is the Debye temperature. We also use the Allen-Dynes\cite{allen} version of the McMillan formula, where the prefactor 
$\frac{\langle\omega_{\rm log}\rangle}{1.20}$ is used, where 
\begin{equation}
\langle\omega_{\rm log}\rangle = \exp\left(\int_0^{\omega_{\mathsf{max}}} F(\omega) \ln\omega\frac{d\omega}{{\omega}} \left/ \int_0^{\omega_{\mathsf{max}}} 
{F(\omega)}\frac{d\omega}{{\omega}} \right. \right).
\end{equation}

As far as the Debye temperature is concerned, there are three values reported in the literature, 310~K (Ref.~\cite{cc-prl}), 283~K (Ref.~\cite{tran-heat}) and 248~K (Ref.~\cite{yan2013}), we take the middle value, $\Theta = 283$~K, as representative, however this particular choice doesn't change the relative difference between two phases. For the modified $T_c$ equation, $\langle\omega_{\rm log}\rangle$ is independently calculated for both phases, and a small difference is found, $\langle\omega_{\rm log}\rangle = 144$~K in tetragonal and $\langle\omega_{\rm log}\rangle = 143$~K in cubic.
Computed $T_c$ is plotted for both phases, as a function of the Coulomb pseudopotential parameter $\mu^{\star}$, in Fig.~\ref{fig-tc-mu} (since the electronic densities of states are very similar in both phases, we expect that the Coulomb repulsion, described by the parameter $\mu^{\star}$, should be practically the same). 
We see, that tetragonal distortion favors superconductivity, since resulting $T_c$ is slightly higher at given $\mu^{\star}$, as a result of higher $\lambda$.
Note, that the numerical prefactor $\frac{\langle\omega_{\rm log}\rangle}{1.20}$ in the Allen-Dynes version of the $T_c$ equation was originally fitted using $\mu^{\star} = 0.1$ instead of McMillan's $\mu^{\star} = 0.13$, thus when Allen-Dynes equation is used, lower $\mu^{\star}$ should be taken. This explains why $T_c$ obtained using Allen-Dynes formula at the same  $\mu^{\star}$ is lower than using McMillan formula.
Nevertheless, the experimental value of $T_c \simeq 2.2$~K is reproduced in calculations for the realistic range of $\mu^{\star} = 0.11 - 0.14$, thus, in the tetragonal phase, the assumption, that the electron-phonon interaction is responsible for the superconductivity in Mo$_3$Sb$_7$ is even better supported, than in the cubic one. 

\section{Magnetism and relativistic effects\label{sec2}}

Results, presented in previous Section, supported the electron-phonon mechanism as being responsible for the superconductivity in Mo$_3$Sb$_7$ by showing, that the electron-phonon interaction was strong enough to drive the system into the superconducting state. It still remains unknown, how the magnetic interactions between Mo atoms in Mo$_3$Sb$_7$ influence the superconducting state. As we mentioned in the Introduction, antiferromagnetic fluctuations were even suggested to be a source of the superconductivity.\cite{tran-pressure}

Investigation of the magnetic properties of Mo$_3$Sb$_7$ using DFT techniques occurred to be difficult, since such effects as local low-dimensional spin singlet dimers\cite{Tra08} or valence bond crystal,\cite{koyama} predicted in Mo$_3$Sb$_7$, are difficult to be taken into account by these methods. 
The only way to mimic a singlet state in typical band-structure calculations is to assume antiparallel ordering of local magnetic moments on nearest atoms and to verify if the self-consistent cycle of calculations will stabilize this state. Several attempts to investigate the possibility of static long-range magnetic ordering in the tetragonal Mo$_3$Sb$_7$ were made.
First, the calculated Stoner parameters on Mo atoms are below the critical 1.0 value, being 0.73 and 0.75 for Mo(8i) and Mo(4e) sublattices, respectively, thus are close to those reported for the cubic phase\cite{bw08} ($\sim 0.7$). In agreement with the Stoner model, spin-polarized calculations for the ferromagnetic spin arrangements yielded non-magnetic ground state. 
As far as anfiterromagnetic configurations are concerned, Koyama {\it et al.}\cite{koyama} proposed the two AFM exchange coupling constants $J_1$ and $J_2$ between Mo(4e)-Mo(4e) and Mo(4e)-Mo(8i). If the local Mo magnetic moments were coupled that way, it could result in the magnetic structure represented by the arrows in the Fig.~\ref{fig-cryst} (i.e. two Mo(4e) atoms are coupled antiferromagnetically to each other and to neighboring four Mo(8i) atoms, with magnetic moments along $z$ direction). For such a case, magnetic unit cell has to be doubled along $c$ axis, and the two Mo atoms, which build the Mo(4e) dimer, have to be considered as nonequivalent, to allow for the opposite direction of the magnetic moment. As a result of distinguished Mo(4e) atoms, the unit cell becomes noncentrosymmetric, as will be discussed below.
For this case, again, the non-magnetic state was predicted by DFT calculations.
All these results were verified using the relativistic full-potential calculations, where the spin-orbit coupling was included, using both LDA and GGA functionals. Thus, any stable, long-range ordering of magnetic moments was not found, in agreement with experimental findings and previous theoretical studies of the tetragonal~\cite{mosb-tetra-calc} and cubic~\cite{bw08} phases. 
Moreover, to verify whether the absence of magnetism in DFT calculations may be related to the arbitrarily assumed type of magnetic ordering, we performed calculations for the so called disordered local moment state (DLM). In this approach, a completely disordered magnetic state, with local non-zero magnetic moments, can be studied, thanks to the using of the coherent potential approximation (CPA) (see, e.g. Refs.\cite{dlm, bw-dlm}). 
Nevertheless, also in these KKR-CPA calculations, all the magnetic moments of the Mo atoms converged to zero values. 

The same set of calculations for the smaller unit cells were carried out, to simulate the effect of external pressure. Also here, both for ordered (FM, AFM) and disordered (DLM) states the local Mo magnetic moments vanished. This proves the subtle nature of the magnetic properties of this system, and is in agreement with the non-magnetic spin-singlet states formation scenario.\cite{Tra08,koyama}

The cubic-to-tetragonal phase transition in  the Mo$_3$Sb$_7$ system is possible under two terms: the dimerisation of two Mo(4e) atoms in a siglet pair\cite{Tra08} and their ordering to the valence bond crystal\cite{koyama} along one of the axis of the cubic structure.
The origin of the second term that guarantees the choice of one axis in singlet formation is not evident and was suggested to be the spin frustration.\cite{koyama} 
As we mentioned in the Introduction, if the 'singlet pair' would consist of two antiparallel individual Mo magnetic moments localized on the two Mo(4e) atoms, rather than being a quantum spin-singlet state, that would result in loosing the center of inversion in such a magnetic unit cell. Since the stabilization of the local magnetic moments, resulting in the antiferromagnetic spin density wave formation was reported under external pressure\cite{tran-pressure}, Mo$_3$Sb$_7$ under pressure may become a member of very interesting group of the noncentrosymmetric superconductors,\cite{bauer-book} but with the noncentrosymmetricity induced by magnetism.

To recall shortly, superconductivity and the presence of the inversion center are connected by the basic symmetry considerations.\cite{anderson59,anderson,sym1,sym2} For the s-wave (BCS-type) pairing, the electrons forming the Cooper pair have opposite spin and momentum, i.e. it is required that when $|{\bf k, \uparrow}\rangle$ state exists near the Fermi level, the state $|-{\bf k, \downarrow}\rangle$ has to exist, to form the Cooper pair. The $|{\bf k}\rangle$, $|-{\bf k\rangle}$ degeneracy is provided by the presence of either inversion center or time reversal symmetry, while the $|{\uparrow}\rangle$, $|{ \downarrow}\rangle$ degeneracy is provided only by the latter. As a result, singlet pairing is not expected in ferromagnetic compounds, due to the time reversal breaking, and triplet pairing is not expected in the noncentrosymmetric structures, due to the lack of the inversion center. Situation becomes more complex, when spin-orbit coupling (SOC) is important in the system, since when SOC is strong, the 
mixing of parity of the superconducting state can be observed.~\cite{sym2,annett}
For example, in the noncentrosymmetric family of Li$_2$Pd$_3$B and Li$_2$Pt$_3$B compounds,\cite{lipdb1,liptb} specific heat~\cite{lipdb-spheat} or NMR~\cite{lipdb-nmr} results support s-wave isotropic superconductivity, in contrast to the penetration depth measurements~\cite{lipdb-penetr}, where mixture of spin-singlet and triplet components in the superconducting energy gap are suggested.

\begin{figure*}[htb]
\includegraphics[width=0.99\textwidth]{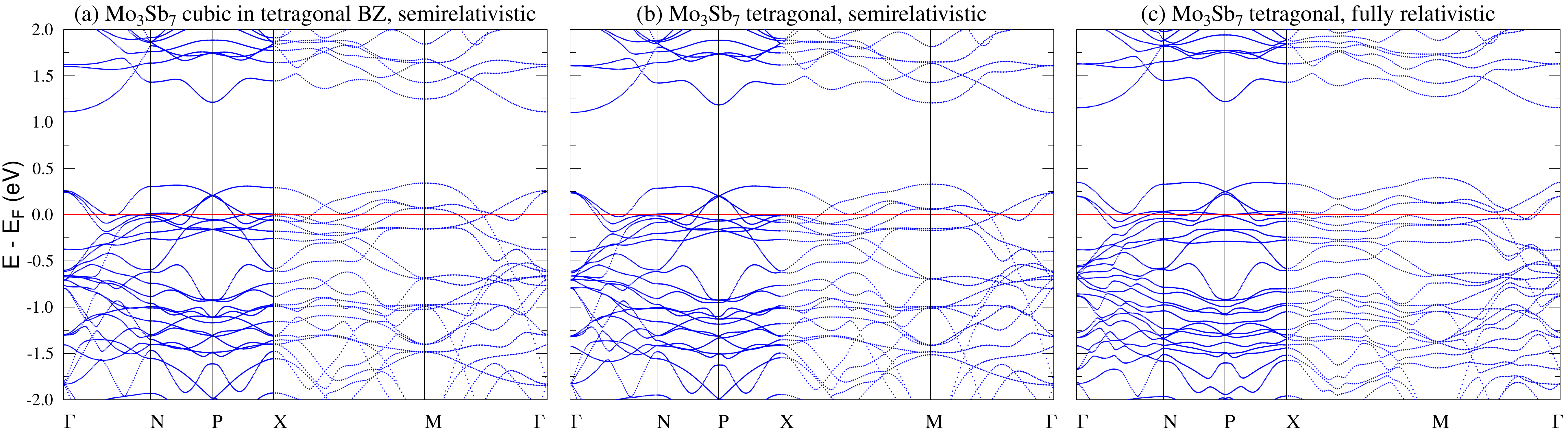}
\caption{\label{fig-bnd}  Electron dispersion relations in the high symmetry directions in Mo$_3$Sb$_7$. Left: cubic phase in the tetragonal Brillouin zone, semi-relativistic calculations, middle: tetragonal phase, semi-relativistic calculations, right: tetragonal phase, fully relativistic calculations.}
\end{figure*}

As the spin-orbit interaction is the important parameter for any noncentrosymmetric effects, relativistic band structure was calculated\cite{PRB2014} for the tetragonal Mo$_3$Sb$_7$. Figure~\ref{fig-bnd} presents the electronic dispersion relations computed in three ways. Fig.~\ref{fig-bnd}(a) shows semirelativistic $E({\bf k})$ relations for the cubic phase, computed in tetragonal symmetry (i.e. we reduce the unit cell symmetry operators to tetragonal but employ the cubic cell and sites parameters). Fig.~\ref{fig-bnd}(b) shows semirelativistic and Fig.~\ref{fig-bnd}(c) fully relativistic $E({\bf k})$, both for the tetragonal phase. 
First we can see, that Fig.~\ref{fig-bnd}(a) and  Fig.~\ref{fig-bnd}(b) are very similar, again confirming the small influence of the distortion on the electronic structure. Differences can be seen e.g. between $X$ and $N$ points, which are equivalent in cubic phase and in-equivalent in tetragonal, also some band degeneracies are removed by distortion at $M$ point.
More visible changes, although still moderate, are induced by the relativistic effects. The largest is the spin-orbit splitting of the highest valence band at $\Gamma$, $\Delta E \simeq 0.15$~eV. Differences are also well seen along the $X$-$M$-$\Gamma$ path, with larger energy levels splitting at $M$ point, than induced by the distortion.
We may conclude, that the spin-orbit interaction has visible effect on the electronic band structure, thus effects of the noncentrosymmetricity (if present) may play a role for the pairing symmetry.

The question, whether any mixing of parity, induced by the lack of inversion center is possible for Mo$_3$Sb$_7$ is certainly related to the question whether any static AFM-like state coexists with the superconductivity.
As was argued by Tran {\it et al.}\cite{tran-pressure} external pressure stabilizes the magnetic state by inducing the spin density wave (SDW) phase below $T_{\rm SDW} = 6.6$~K at p = 4.5 kbar. Both, SDW and superconducting phases, although competes with each other ($T_c$ increases with p, $T_{\rm SDW}$ decreases for p $>$ 4.5 kbar), coexist in some pressure ranges. 
If so, superconductivity may be present in parallel with the noncentrosymmetricity, induced by SDW phase. Repetition of the basic experiments (e.g. specific heat, NMR or $\mu$SR) to reveal the gap symmetry under pressure should shed some light on the pairing symmetry.
Especially, that Andreev reflexion studies\cite{andreev1,andreev2} in Mo$_3$Sb$_7$ even at ambient pressure suggested non-BCS strong gap anisotropy, with possible s+g gap symmetry.

\section{Summary}\label{sec3}
Electronic and dynamical properties of the tetragonal phase of Mo$_3$Sb$_7$ were analyzed, using first principles density functional methods. Electron-phonon coupling constant was estimated within the rigid muffin tin approximation, without any adjustable parameters. We found, that the tetragonal distortion has small and beneficial effect on superconductivity, slightly increasing the coupling constant, from $\lambda_{\rm cub} = 0.54$ to $\lambda_{\rm tet} = 0.55$. Effect on the superconducting critical temperature $T_c$ was evaluated using McMillan-type formulas, and calculated $T_c$'s for the tetragonal phase were also slightly higher than corresponding values for the cubic one. The agreement of the calculated $T_c$ with the experimental values supports the electron-phonon interaction as the pairing mechanism. 
The spin-polarized calculations for the ordered (ferromagnetic or antiferromagnetic), as well as disordered (DLM) magnetic states yielded non-magnetic ground states.
Next, we pointed out, that if Mo(4e) atoms are physically nonequivalent, as may be suggested by the formation of SDW phase under pressure, Mo$_3$Sb$_7$ structure may be regarded as noncentrosymmetric. In this context, relativistic band structure was calculated and spin-orbit interaction was found to have a visible influence on the electronic bands near the Fermi level. Thus, mixing of the superconducting pairing symmetries might be considered for the description of the pressure experiments, where coexistence of spin density wave and superconductivity was reported.~\cite{tran-pressure}

\section*{Acknowledgments}

BW was partly supported by the Polish National Science Center (project no. DEC-2011/02/A/ST3/00124) and the Polish Ministry of Science and Higher Education.


\section*{References}

\begin{thebibliography}{99}

\bibitem{Buk02} Z. Bukowski, D. Badurski, J. Stepien-Damm,
                   and R. Tro\'c,
                   Solid State Commun. {123} (2002) 283.

\bibitem{Buk06} V. M. Dmitriev, L. F. Rybaltchenko, L. A. Ishchenko,
                   E. V. Khristenko, Z. Bukowski, and R. Tro\'c,
                   Supercond. Sci. Technol. {19} (2006) 573.

\bibitem{cc-prl} C. Candolfi, B. Lenoir, A. Dauscher, C.
                   Bellouard, J. Hejtmanek, E. Santava, and J. Tobola,
                   Phys. Rev. Lett. {99} (2007) 037006 .

\bibitem{Tra08} V. H. Tran, W. Miiller, and Z. Bukowski,
                   Phys. Rev. Lett. {100} (2008) 137004.

                   
\bibitem{tran-gap-evidence} V. H. Tran, A. D. Hillier, D. T. Adroja, Z. Bukowski, and W. Miiller, J. Phys.: Condens. Matter, {21} (2009) 485701.

                   
\bibitem{koyama}  T. Koyama, H. Yamashita, Y. Takahashi, T. Kohara, I. Watanabe, Y. Tabata, and H. Nakamura, Phys. Rev. Lett. {101} (2008) 126404.
                   

\bibitem{okabe} H. Okabe, S. Yano, T. Muranaka, and J. Akimitsu, J. Phys.:
Conference Series {150} (2009) 052196.

\bibitem{koyama2} T. Koyama, H. Yamashita, T. Kohara, Y. Tabata, and
H. Nakamura, Mater. Res. Bull. {44} (2009) 1132.
                                      
                   
\bibitem{tran-heat} V.~H. Tran, W. Miiller, and Z. Bukowski,
                 Acta Mater. {56} (2008) 5694.

                   
\bibitem{cc-heat} C. Candolfi, B. Lenoir, A. Dauscher, J. Hejtmanek,
                   E. Santava, and J. Tobola,
                   Phys. Rev. B {77} (2008) 092509.


\bibitem{bw08}  B. Wiendlocha, J. Tobola, M. Sternik, S. Kaprzyk, K. Parlinski, and A. M. Ole\'s, Phys. Rev. B {78} (2008) 060507(R).

                   
\bibitem{yan2013} J.Q. Yan, M. A. McGuire, A. F. May, H. Cao, A. D. Christianson, D. G. Mandrus, and B. C. Sales
                   Phys. Rev. B {87} (2013) 104515.


\bibitem{usr-khasanov} R. Khasanov, P. W. Klamut, A. Shengelaya, Z. Bukowski, I. M. Savi\'c, C. Baines, and H. Keller, Phys. Rev. B {78} (2008) 014502.

\bibitem{usr-tran} V. H. Tran, A. D. Hillier, D. T. Adroja, and Z. Bukowski, Phys. Rev. B {78} (2008) 172505. 
                   
                   
\bibitem{tran-pressure} V. H. Tran, R. T. Khan, P. Wi\'sniewski, and E. Bauer, J. Phys.: Conf. Ser. {273} (2011) 012088.

\bibitem{mosb-transport} C. Candolfi, B. Lenoir, A. Dauscher, E. Guilmeau, J. Hejtmanek, J. Tobola, B. Wiendlocha, and S. Kaprzyk, Phys. Rev. B {79} (2009) 035114 .

\bibitem{mosb-thermo1} C. Candolfi, B. Lenoir, A. Dauscher, J. Hejtmanek, and J. Tobola, Phys. Rev. B {79} (2009) 235108.


\bibitem{mosb-thermo2} C. Candolfi, B. Lenoir, A. Dauscher, J. Hejtmanek, and J. Tobola, Phys. Rev. B {80} (2009) 155127.


\bibitem{mosb-thermo3} C. Candolfi, B. Lenoir, C. Chubilleau, A. Dauscher, and E. Guilmeau, J. Phys.: Condens. Matter {22} (2010) 025801.


\bibitem{mosb-thermo4} X. Shi, Y. Pei, G. J. Snyder, and L. Chen, Energy Environ. Sci. {4} (2011) 4086.


\bibitem{mosb-thermo5} D. Parker, M-H Du, and D. J. Singh, Phys. Rev. B {83} (2011) 245111.

\bibitem{mosb-thermo6} X. Deng, Minfeng L\"u, J. Meng, J. Alloys Compd. {577} (2013) 183.

\bibitem{mosb-tetra-calc} S. Nazir, S. Auluck, J. J. Pulikkotil, N. Singh, and U. Schwingenschlogl, Chem. Phys. Lett. 504 (2011) 148.
                   

\bibitem{mcm}   W. L. McMillan,
                   Phys. Rev. {167} (1968) 331.


\bibitem{hop}   J. J. Hopfield,
                   Phys. Rev. {186} (1969) 443.

                   
\bibitem{allen} P. B. Allen and R. C. Dynes,
                   Phys. Rev. B {12} (1975) 905.


\bibitem{kkr99} A. Bansil, S. Kaprzyk, P. E. Mijnarends, and J. Tobola,
                   Phys. Rev. B {60} (1999) 13396.

\bibitem{lda}   U. von~Barth and L. Hedin,
                   J. Phys.: Condens. Matter {5} (1972) 1629.


\bibitem{Par97} K. Parlinski, Z-. Q. Li, and Y. Kawazoe,
                   Phys. Rev. Lett. {78} (1997) 4063 ;
                K. Parlinski,
                   Computer code {\sc phonon}, Cracow, 2008.

\bibitem{vasp}  G. Kresse and J. Furthm\"uller,
                   Comput. Mater. Sci. {6} (1996) 15;
                   Phys. Rev. B {54} (1996) 11169.

\bibitem{Per96} J. P. Perdew, K. Burke, and M. Ernzerhof,
                   Phys. Rev. Lett. {77} (1996) 3865.

\bibitem{GenPDOS} C. Candolfi, B. Lenoir, A. Dauscher, 
                  M. M. Koza, M. de Boissieu, M. Sternik and K. Parlinski, 
                    Phys. Rev. B {84} (2011) 224306.

\bibitem{rmta}  G. D. Gaspari and B. L. Gy\"orffy,
                   Phys. Rev. Lett. {28} (1972) 801 ;
                I. R. Gomersall and B. L. Gy\"orffy,
                   J. Phys. F {4} (1974) 1204.

\bibitem{pickett} W. E. Pickett,
                   Phys. Rev. B {25} (1982) 745.

\bibitem{prb-bw} B. Wiendlocha, J. Tobola, and S. Kaprzyk,
                   Phys. Rev. B {73} (2006) 134522.

\bibitem{pss-bw} B. Wiendlocha, J. Tobola, S. Kaprzyk, D. Fruchart, and J. Marcus,
		  Phys. Stat. Sol. B, {243} (2006) 351. 
                   
\bibitem{cr3gan-bw} B. Wiendlocha, J. Tobola, S. Kaprzyk, and D. Fruchart,
		    J. Alloys Compd,  {422} (2007) 289.

\bibitem{bw-dlm} B. Wiendlocha, J. Tobola, S. Kaprzyk, R. Zach, E. K. Hlil, and D. Fruchart,
		    J. Phys. D: Appl. Phys. {41} (2008) 205007,

\bibitem{dlm} A. J. Pindor, J. Staunton, G. M. Stocks, and H. Winter, J. Phys. F: Met. Phys. {13} (1983) 979.
        
               
\bibitem{grimvall} G. Grimvall, {\it The electron-phonon interaction in metals} (North-Holland, Amsterdam, 1981).

\bibitem{bauer-book} E. Bauer and M. Sigrist (Eds.), {\it Non-Centrosymmetric Superconductors} (Springer, Heidelberg, 2012).

\bibitem{anderson59} P. W. Anderson, J. Phys. Chem Solids, {11} (1959) 26.

\bibitem{anderson} P. W. Anderson, Phys. Rev. B {30} (1984) 4000.

\bibitem{sym1} P. A. Frigeri, D. F. Agterberg, A. Koga, and M. Sigrist, Phys. Rev. Lett., {92} (2004) 097001.

\bibitem{sym2} K. V. Samokhin, E. S. Zijlstra, and S. K. Bose,  Phys. Rev. B {69} (2004) 094514.

\bibitem{annett} J. Quintanilla, A. D. Hillier, J. F. Annett, and R. Cywinski, Phys. Rev. B {82} (2010) 174511.

\bibitem{lipdb1} K. Togano, P. Badica, Y. Nakamori, S. Orimo, H. Takeya, and K. Hirata, Phys. Rev. Lett. {93} (2004) 247004.

\bibitem{liptb} P. Badica, T. Kondo, and K. Togano, J. Phys. Soc. Japan {74} (2005) 1014.

\bibitem{lipdb-spheat} H. Takeya, K. Hirata, K. Yamaura, K. Togano, M. El Massalami,R. Rapp, F. A. Chaves, and B. Ouladdiaf Phys. Rev. B {72} (2005) 104506.

\bibitem{lipdb-nmr} M. Nishiyama, Y. Inada, and G. Q. Zheng Phys. Rev. B {71} (2005) 220505(R).

\bibitem{lipdb-penetr} H. Q. Yuan, D. F. Agterberg, N. Hayashi, P. Badica, D. Vandervelde, K. Togano, M. Sigrist, and M. B. Salamon Phys. Rev. Lett. {97} (2006) 017006.

\bibitem{PRB2014} K. Kutorasinski, B. Wiendlocha, J. Tobola, and S. Kaprzyk, Phys. Rev. B 89 (2014) 115205.

\bibitem{andreev1} V. M. Dmitriev, L. F. Rybaltchenko, L. A. Ishchenko, E. V. Khristenko, Z. Bukowski, and R. Tro\'c, Low Temp. Phys. {33} (2007) 1009.

\bibitem{andreev2} V. M. Dmitriev, L. F. Rybaltchenko, E. V. Khristenko, L. A. Ishchenko, Z. Bukowski, and R. Tro\'c, Low Temp. Phys. {33} (2007) 295.

\end{thebibliography}
\end{document}